\documentclass[12pt,preprint]{aastex}

\slugcomment{Submitted to ApJ}

\shorttitle{Close companions of the BL Lac Objects 0502+675 and 1440+122}
\shortauthors{Giovannini et al.}

\def\ref{\par \noindent \hang}

\def\mincir{\ \raise -2.truept\hbox{\rlap{\hbox{$\sim$}}\raise5.truept
\hbox{$<$}\ }}

\def\ltsim{\raise 2pt \hbox {$<$} \kern-1.1em \lower 4pt \hbox {$\sim$}}
\def\gtsim{\raise 2pt \hbox {$>$} \kern-1.1em \lower 4pt \hbox {$\sim$}}

\begin{document}

\title{The nature of close companions of the BL Lac Objects 1ES~0502+675 and 
1ES~1440+122}

\author{G. Giovannini$^1$}
\affil{Dipartimento di Astronomia, dell'Universita' di Bologna, via Ranzani 1, 
40127 Bologna, Italy}
\altaffiltext{1}{
Istituto di Radioastronomia del CNR, via Gobetti 101, 40129 
Bologna, Italy}
\email{ggiovannini@ira.cnr.it}

\author{R. Falomo}
\affil{INAF -- Osservatorio Astronomico di Padova, Vicolo dell'Osservatorio 5, 
35122 Padova, Italy}
\email{falomo@pd.astro.it}

\author{R. Scarpa}
\affil{European Southern Observatory, 3107 Alonso de Cordova,
Santiago, Chile.}
\email{rscarpa@eso.org}

\author{A. Treves}
\affil{Dipartimento di Fisica e Matematica, dell'Universit\`a dell'Insubria, 
via Valleggio 11, 22100 Como, Italy}
\email{treves@mib.infn.it}
\and
\author{C. M. Urry}
\affil{Department of Physics, Yale University, PO Box 208121, New Haven,
CT 06520-8121, USA}
\email{meg.urry@yale.edu}

\begin{abstract} 
We report on deep radio images and optical spectroscopy of two 
BL Lac objects that have very close compact companions.
The two targets, 1ES 0502+675 and 1ES 1440+122, 
were selected from the HST imaging survey of 110 BL Lacs 
as candidates for possible gravitational lensing. 
The new observations clearly demonstrate that the companion objects are not 
secondary images of the active nuclei but, in spite of the relatively low 
chance projection probability, foreground Galactic stars. 
Gravitational lensing appears to be unimportant to the BL Lac phenomenon.
We discuss the radio properties of the BL Lac objects in 
the context of standard beaming models, and show they are
as expected for beamed FR~I radio galaxies.
\end{abstract}

\keywords{BL Lacertae objects: general -- galaxies:active -- 
galaxies: elliptical and lenticular -- galaxies: kinematics and dynamics -- 
galaxies:nuclei}

\vfill\eject

\section{Introduction}

BL Lac objects and other blazars are characterized by their
rapid large-amplitude continuum variability, which is usually 
explained by relativistic beaming in a jet emanating from 
the active galactic nucleus \citep{ur95}.
However, \citet{os85, os90} pointed out that gravitational 
micro-lensing of a compact continuum source by stars in the core
of an intervening galaxy
could enhance this continuum variability, and furthermore,
could explain the peculiar defining properties of BL Lac objects, 
namely, weak or absent emission lines. 
According to this idea, gravitational micro-lensing magnifies 
the nuclear light from the background quasar but not that 
of the larger emission-line region, thus diluting the lines
and reducing their equivalent widths.

If the gravitational lensing hypothesis is correct it implies 
at least four observable effects.
First, the nuclei of BL Lac objects should be slightly off-centered 
with respect to the surrounding nebulosity (which would be the
intervening galaxy rather than a host galaxy).
Second, the nebulosity is about 20 times more likely to be
an elliptical galaxy than a spiral galaxy, because of the
relation between velocity dispersion and lensing probability
\citep{os90}.
Third, an excess of intervening absorption systems should be 
observed in BL Lacs spectra.
Fourth, because of the macro-lensing by the entire intervening galaxy,
in some cases multiple images of the active nucleus should be observed
(as in the case of QSO lenses).

Observations of some individual sources in the past decade 
have offered mixed support for the gravitational lensing hypothesis.
On the one hand, it appeared in some cases that the 
surrounding nebulosity was off-center with respect to the BL Lac 
nucleus (e.g., AO~0235+164: \citet{st88a,ab93};
PKS~0537-441: \citet{st88b,le00}; 
MS~0205.7+3509: \citet{st95,fa97}).
However, because of the difficulty of such observations 
(in terms of spatial resolution, background subtraction 
and appropriate definition of the PSF) most of these results were not 
confirmed by subsequent studies (\citet{fa96,fa92,fa97,he03,pi02,
ni96,pu02,sc00}).

At the same time, Mg~II absorption systems are significantly
more common in BL Lac spectra than in quasar spectra
\citep{st97}, suggesting an unusually high 
occurrence of foreground objects. 
However, VLA radio images of these same BL Lac objects 
do not show the expected gravitational lensing features 
\citep{re03}, so again,
the relevance of gravitational lensing to the BL Lac phenomenon
is controversial.

The HST\footnote{Based on observations made with the
NASA/ESA Hubble Space Telescope, obtained at the Space Telescope Science
Institute, which is operated by the Association of Universities for
Research in Astronomy, Inc., under NASA contract NAS~5-26555.}
snapshot survey of BL Lac objects 
(\citet{ur00,sc00}) 
provided a homogeneous set of 110 high-resolution 
images through the F702W filter (approximately R-band),
analysis of which yielded a full characterization 
of the host galaxies and a detailed view of their close environments.
In particular, for low redshift ($z < 0.3$) 
\citet{fa00} showed that the 
nuclei of BL Lacs are centered (within an accuracy of 0.05 arcsec) 
with respect to their surrounding nebulosity, 
contrary to what is predicted by the gravitational lensing hypothesis.
\citet{ur00} also found that the host galaxy 
morphologies were overwhelmingly elliptical rather than disky,
perhaps more than lensing would predict.

Yet the HST snapshot survey also yielded some possible evidence 
in favor of gravitational lensing.
In particular, \citet{sc99} reported a number of 
close compact companions, which, based on the limited HST
photometry available, were consistent with lensed images.
Lensing had been seen previously in HST images of quasars
\citep{ma93}, albeit at a much lower rate (1 lens
in $\sim500$ quasar images, compared to 3 lens candidates
in 110 BL Lac images).

The mixed evidence, and the lingering possibility that 
lensing is important in BL Lac objects, make it essential 
to follow up on the three candidate lenses from the 
HST snapshot survey \citep{sc99}. 
One candidate was a possible arc or Einstein ring 
which will not be discussed here (O'Dowd et al. in preparation).
The other two candidates,
1ES 0502+675 ($z=0.314$) and 1ES 1440+122 ($z=0.162$), 
showed possible multiple images with 
separations of $\lesssim 0.3$~arcsec. 
In the case of 1ES 0502+675, the case was strengthened 
by an HST NICMOS H-band image that showed 
the colors of the nucleus and the companion were similar \citep{sc99}. 
However the lack of further photometric information or spectroscopy 
of the candidate lenses prevented any definitive conclusion.

We report here the results of two specific 
observational programs carried out in order to clarify the
nature of these two BL Lac objects.
The first program, aimed at obtaining deep VLA images of the two
fields, is described in \S~2. 
The second program, described in \S~3, 
involved HST optical spectroscopy of the two BL Lacs 
and their companions.
The results of both programs, which indicate that the companions are 
foreground Galactic stars, are given in \S~4, along with a brief 
discussion of the radio properties of the two BL Lac objects. 
The main conclusions from this work are summarized in \S~5.
In this paper we use H$_0$ = 70 km sec$^{-1}$ Mpc$^{-1}$ and
q$_0$ = 0.5. 

\section{Radio Observations and data analysis}

The two BL Lac objects 1ES 0502+675 and 1ES 1440+122 
were observed with the NRAO VLA\footnote{The
National Radio Astronomy Observatory is a facility 
of the National Science Foundation operated under 
cooperative agreement by Associated Universities, Inc.}
in the A configuration on
1999 June 21 (1ES 0502+675) and June 24 (1ES 1440+122) for
2 hours each at 8.4, 15, and 22~GHz (see Tables 1 and 2).
Both objects were observed again
several months later (1ES 1440+122, 11 November; 1ES 0502+675, 14 November),
with the VLA in the B configuration in order to have better
uv coverage to search for a possible extended radio emission. 
To maximize the sensitivity to faint features,
the second-epoch observations of both sources were 
done only at 8.4~GHz.
The data were calibrated in the standard way using the NRAO AIPS package 
and reduced using the AIPS task IMAGR.
Calibrated data at 8.4~GHz in the A and B configurations were combined 
to obtain a more sensitive map. 
The angular resolution and sensitivity of the final calibrated
images are given in Tables 1 and 2.

\subsection{Radio images}

{\bf 1ES 0502+675} 

The VLA images show that this BL Lac object is 
unresolved in the A configuration at all frequencies and in the 
longer observation 
in the B configuration at 8.4~GHz. 
In the latter observation the source flux density 
is consistent with the flux density at the same frequency
measured at higher resolution a few months before. This confirms that
no extended low surface brightness structure is present and that no flux
density variability occurred between the two observations.
The 15~GHz image of 1ES 0502+675 is shown in Fig.~1.

In Table 1 we report the relevant parameters from the new images together 
with the data from the NVSS survey at 1.4~GHz \citep{co98}. 
The spectral index (S($\nu) \propto \nu^{-\alpha}$) is flat with a possible
indication of self-absorption at $\sim 8.4$~GHz: 
$\alpha_{1.4}^{8.4} = -0.11$ and
$\alpha_{8.4}^{22.5} = 0.36$. 
The low-resolution (HPBW = $45^{\prime\prime}$) 
1.4~GHz flux density confirms the lack of any extended structure.

\medskip
{\bf 1ES 1440+122} 

VLA images in the A configuration show only the 
unresolved nuclear emission from this BL Lac object
(Fig.~2 and Table~2). The radio spectrum is flat,
$\alpha_{8.4}^{22.5} = 0.16$, 
but at 8.4~GHz the total flux in the shortest baselines is 
significantly higher than the peak flux density. In the
observation in the B configuration, 
thanks to the presence of short baselines
and to the longer integration time, 
extended low-brightness emission is 
visible around the nuclear source. 

In Fig.~3 we show the image obtained
by combining the A and B configuration data at 8.4~GHz. 
No evidence of a variation in the core flux density is present 
between the two observations 5 months apart. In the 1.4~GHz image 
from the NVSS survey \citep{co98} the source is only marginally 
resolved because of the large beam. The spectrum is steep at 
low frequency with an evident flattening at high frequency, 
as is typical of BL Lac objects in which a flat-spectrum core dominates 
at high frequencies and an extended steep-spectrum component dominates 
at low frequencies.

\subsection{Comparison with HST optical images}

HST WFPC2 images of 1ES 0502+675 and 1ES 1440+122 in the F702W filter 
are shown in Figs.~1 and 2 (gray scale) together with 
radio maps at 15~GHz (contours). 
The optical exposures were 740 and 320 seconds, respectively. 
Details of the analysis procedures can be found in 
\citet{sc99,sc00,ur00,fa00}. 

For 1ES 0502+675, the magnitude of the brightest source is 
m$_{\rm R}=17.3 \pm 0.1$~mag, and that of the companion is
m$_{\rm R}=18.7 \pm 0.2$~mag. 
For 1ES 1440+122, the central source is 
m$_{\rm R}=16.9 \pm 0.1$~mag, and the closest companion
has m$_{\rm R}=19.8 \pm 0.2$~mag.
Note that the companion galaxy $\sim 2$~arcsec NW of 
1ES 1440+122 does not emit at radio frequencies.

Comparing the images and the radio and optical positions
it is clear that in both cases, 
the radio emission is identified with the BL-Lac object
at the center of a giant elliptical galaxy. 
We note that the radio images were self-calibrated and we estimate 
an astrometric precision of $\sim 0.1^{\prime\prime}$,
and the absolute accuracy of HST astrometry is comparable.
However, 
considering 
that the two secondary objects have been identified
as local stars (see \S~3), and 
that all BL Lac objects are radio loud,
we are confident in our identification of the radio sources 
with the BL Lac objects.

\section{Optical spectroscopy of close companions}

Optical spectra of the two BL Lac objects and their companions
were obtained with the Space Telescope Imaging Spectrograph 
(STIS) using two different setups: 
grism G750L to cover the region around 7500 \AA\ and 
grism G430L to cover the region around 4300 \AA. 
In both cases a slit of $52 \times 0.2$~arcsec 
was centered on the companion and oriented in the 
direction of the bright BL Lac nucleus.
The observations took place on March 9th 2001 for 1ES 0502+675 and
on December 22nd 2002 for 1ES 1440+122.
The total integration times for each spectrum were 2300 seconds for 
1ES 0502+675
and 2000 seconds for 1ES 1440+122.
Standard data reduction was performed using the HST pipeline 
to produce two-dimensional wavelength- and flux-calibrated spectra.
Extraction of 1-dimensional spectra was then performed 
using the standard routines available in IRAF.

The extracted spectra of the two close companions 
are shown in Figures 4 and 5.
Because of the narrowness of the slit and the high level of cosmic rays 
the extracted spectra are somewhat noisy, but they are quite
obviously not typical of BL Lacs, which are dominated by
featureless power-law continua. In low-redshift BL Lacs such
as these two objects, there are sometimes stellar absorption
features arising in the host galaxy, but the features seen here
are clearly at zero redshift.
For the companion of 1ES 0502+675 the detected absorption features
indicate that it is a star of intermediate spectral type (likely F or G).
In the case of 1ES 1440+122 no clear absorptions are
observable but the shape of the continuum suggests a later spectral type star 
(likely K).

\section{Discussion}

\subsection{Testing the lensing hypothesis}

We have tested the hypothesis that unresolved companions 
$\sim0.3^{\prime\prime}$ from two  BL Lac objects
be lensed images of the nuclei. Both deep radio 
observations of the fields and optical spectroscopy 
of the companions yield a negative result.
The companions are not emitting at radio frequencies 
(in sharp contrast to the BL Lac nuclei) and 
their optical spectra exhibit continua and (for 1ES 0502+675) 
absorption features typical of intermediate spectral type stars.
Therefore, despite the relatively low probability of having 
such close projections by chance given the observed 
surface density of stars at intermediate Galactic latitudes,
the observed close companions are indeed stars unrelated 
to the BL Lac sources.

From a dataset of 30 HST fields centered on BL Lacs at $z < 0.3$,
we estimated \citep{fa00} a probability P=0.06 
of having one or more moderately bright companions 
within a radius of 0.5~arcsec of any position.
For the whole dataset (110 images) of the HST snapshot survey,
assuming the same average surface brightness for Galactic stars,
one would expect to observe $\sim 0.05$ chance projections 
with separation $d < 0.3$~arcsec in the whole survey, while 2 
were observed. Thus this discrepancy is just an unlikely
statistical fluctuation.

\subsection{The radio properties of 1ES 0502+675 and 1ES 1440+122}

{\bf 1ES 0502+675}: In the radio band this source appears as a 
classical BL Lac-type object dominated by nuclear emission 
with a flat spectrum.
Since no extended emission, such as a resolved jet,
was detected in this source,
we compare its radio properties with those of FR~I radio galaxies
\citep{gi01} assuming a parsec-scale jet velocity with
Lorentz factor $\gamma = 10$ oriented at $\theta = 5^\circ$ with respect
to the line of sight. From the observed core radio power at
5~GHz, $P_5=10^{24.8}$~W/Hz, we infer an intrinsic core radio power
$P_i = P_{observed}/\delta^2$ $= 10^{22.7}$~W/Hz assuming a spectral 
index $\alpha = 0$.

We use this estimate and the general correlation between core and
total radio power \citep{gi01} to derive the expected
unboosted total flux at 408~MHz,
$P_{tot}(408) = 10^{23.2}$~W/Hz.
At the source redshift ($z= 0.314$), this corresponds to
a rather modest total flux density, 0.8~mJy, in agreement 
with the non-detection of any extended flux density 
in the 1.4~GHz NVSS data.

With its low intrinsic radio power, 1ES 0502+675 is very
comparable to nearby low-power FR~I radio galaxies,
such as the B2 radio galaxies. It is visible at 
$z = 0.314$ only because of relativistic beaming,
illustrating how relativistic jets are present
also in low-power radio sources.

{\bf 1ES 1440+122}: For this BL Lac object we detected 
in the radio band both the extended and the core emission. 
The observed nuclear radio power is $P_c(5.0) = 10^{24.2}$~W/Hz 
and the estimated radio power at 408~MHz, derived from the 
measured extended emission assuming $\alpha = 0.7$, is 
$P_{tot}(408) = 10^{25.0}$~W/Hz. From the core dominance 
\citep{gi01} we derive that the jet
velocity has to be $v/c > 0.84$ and 
the orientation angle $\theta \lesssim 30^\circ$. 
Assuming a fast jet with $\gamma \sim 10$, the
orientation angle is roughly $\theta \sim 28-30$ degrees. 
This source thus appears similar to high-intermediate 
radio power FR~I radio galaxies and its de-projected 
linear size should be $\sim 50$~kpc in agreement
with the size-radio power distribution for 
B2 radio galaxies (\citet{ru90} and Figure 11 in \citet{la04}).

\section{Conclusions}

The main result of this work is that the companions detected 
within 0.3 arcseconds of each of the BL Lac objects studied,
1ES 0502+675 and 1ES 1440+122, are not lensed images. 
This conclusion is supported by both optical spectroscopy and radio 
observations of the two fields.
The companions are galactic stars seen by chance
in close alignment with the BL Lac nuclei.
In these two BL Lac objects --- the only lens candidates 
in the large HST snapshot survey --- gravitational lensing 
does not occur.
Thus the present observations lead further support against the idea that
gravitational microlensing is needed,
or even relevant, to explain the peculiar properties of 
any BL Lac objects. 

Apart from lensing the new radio observations also
allow us to characterize the radio properties of the BL Lac
objects themselves. These are well consistent with the 
the average properties of the class, seen at the
redshift of the two objects, and with being beamed versions
of FR~I galaxies.

\acknowledgments

We thank the Referee for useful comments.
This work has received partial support under contracts
ASI I/R/073/02 and I/R/086/02, and
NASA grant NAG5-9327.


\begin{figure}
\plotone{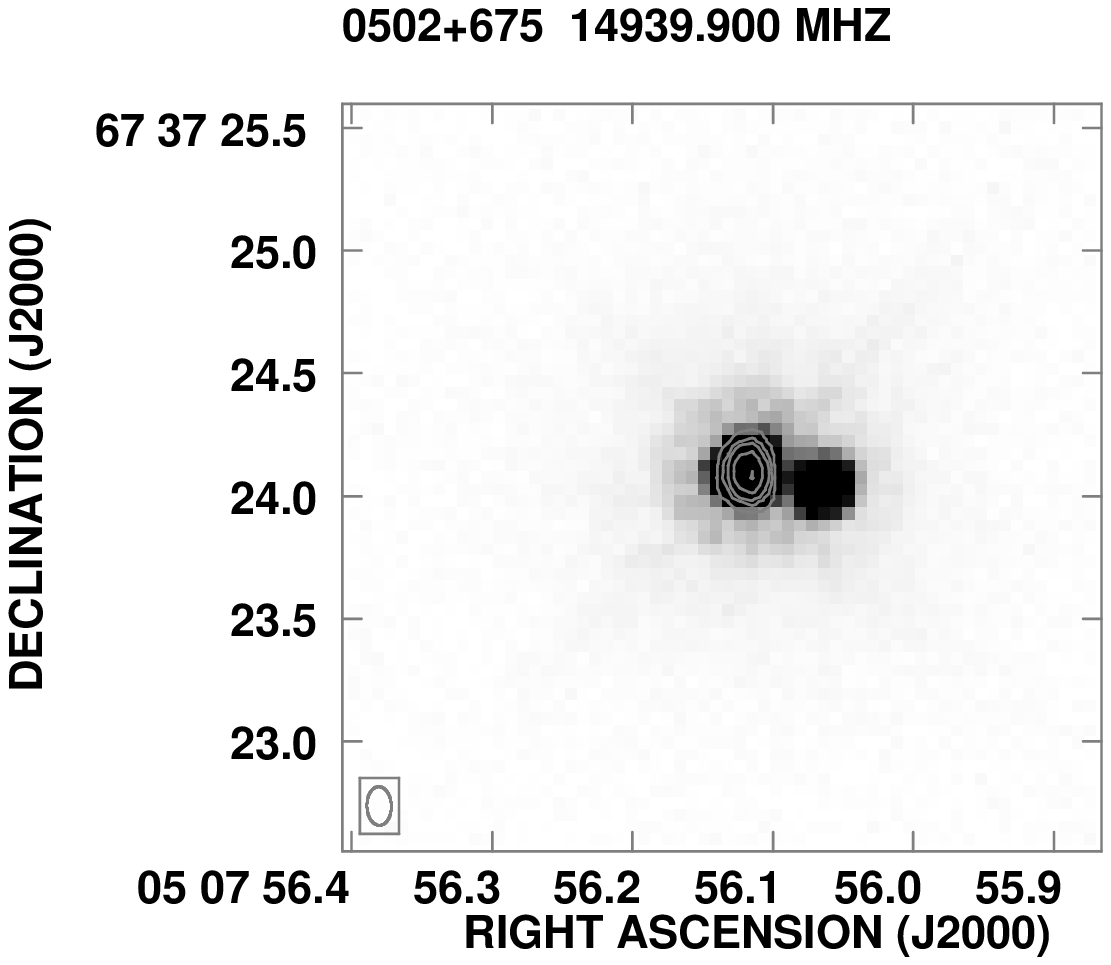}
\caption{VLA contour map of 1ES 0502+675 at 15~GHz superimposed on
the optical HST image in gray scale. While the HST image clearly
shows two point-like sources, only one is detected with the VLA,
indicating very different spectral energy distributions that are
incompatible with gravitational lensing.}
\label{fig1}
\end{figure}

\begin{figure}
\plotone{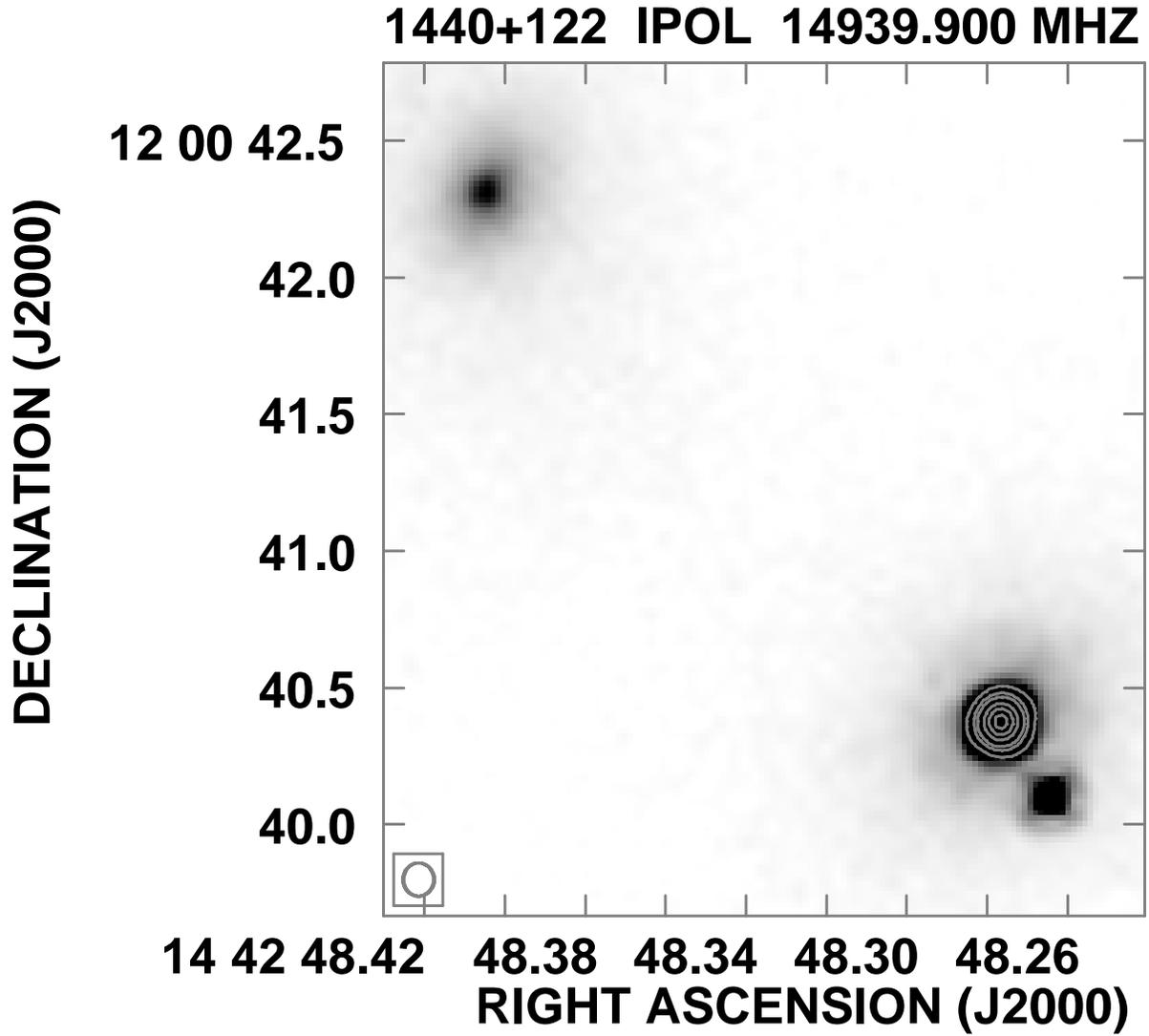}
\caption{VLA contour map of 1ES 1440+122 at 8.4~GHz superimposed 
on the optical HST image in gray scale. The HST image shows
three co-linear sources, only one of which is detected with
the VLA, suggesting the two outlying images are not lensed images.}
\label{fig2}
\end{figure}

\begin{figure}
\plotone{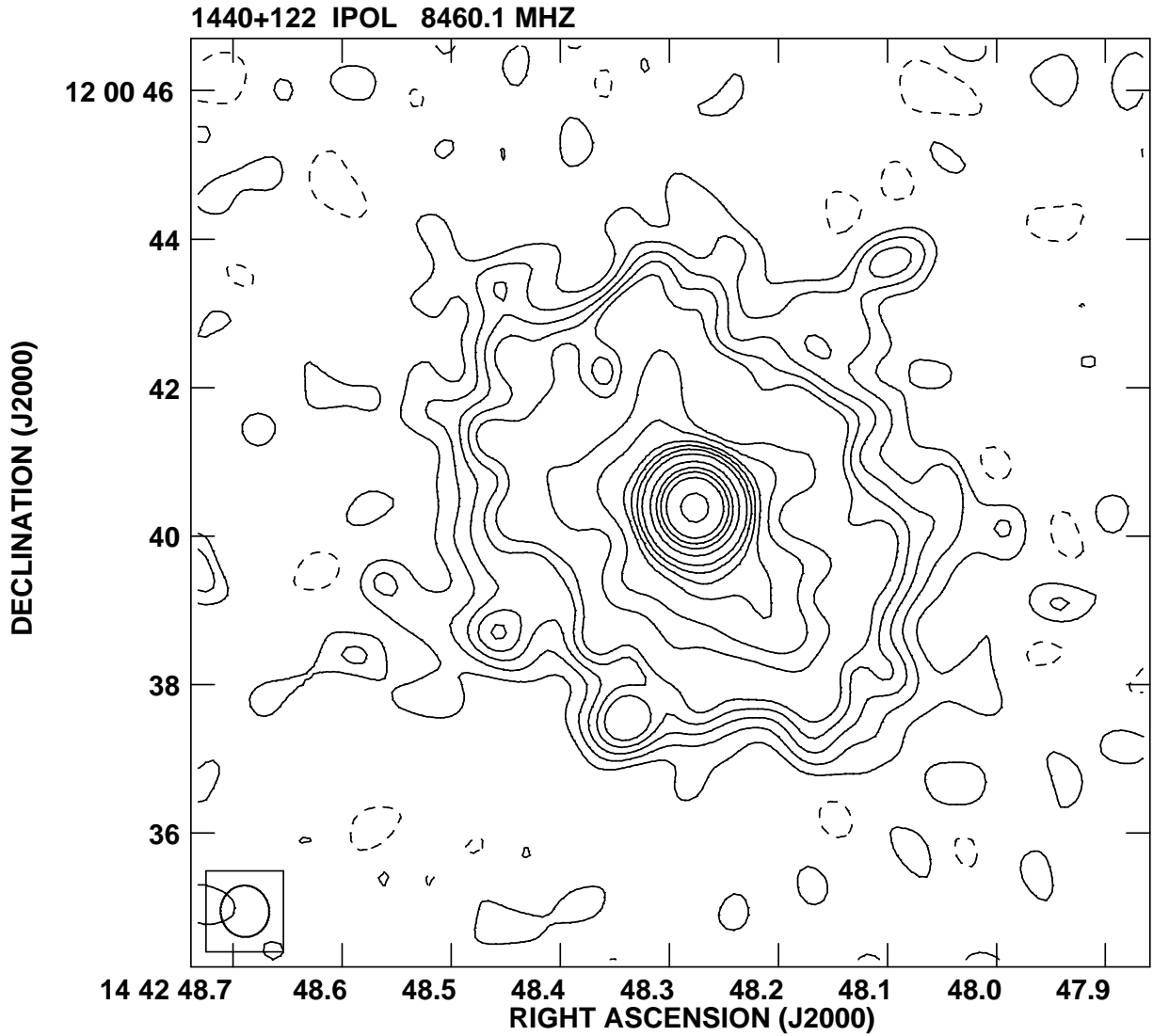}
\caption{8.4~GHz VLA image of 1ES 1440+122 from the combined
A+B configuration data. Contour levels are: -0.03, 0.02, 0.04, 0.06, 0.08, 
0.1, 0.2, 0.3, 0.5, 0.7, 1, 1.5, 3, 5, 7, 10, 20 mJy/beam. The HPBW is 
0.61''$\times$0.59'' in PA 4$^\circ$.}
\label{fig3}
\end{figure}

\begin{figure}
\plotone{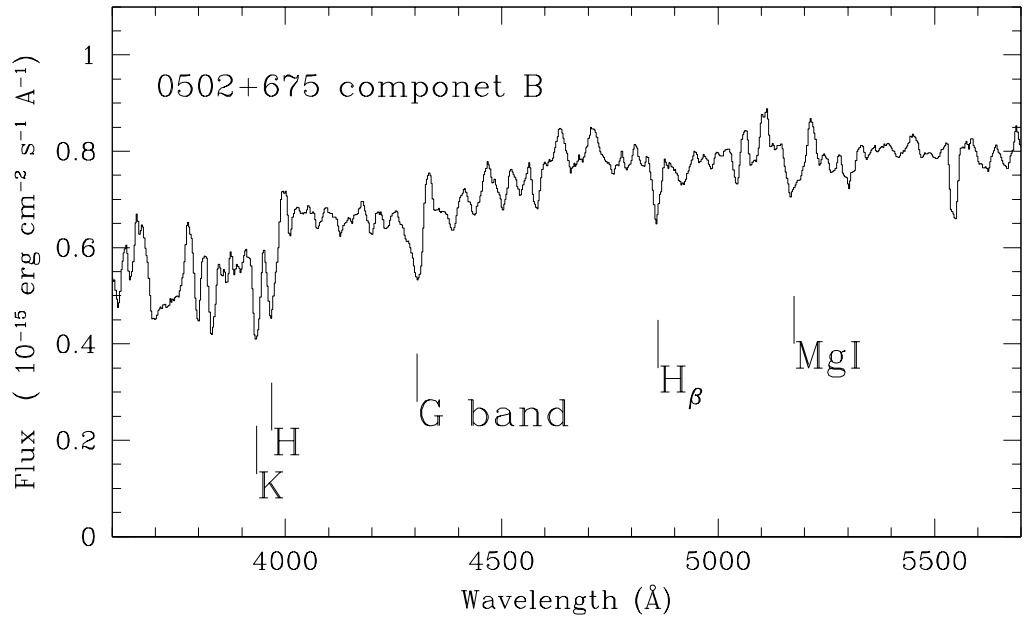}
\caption{STIS optical spectrum of the close companion 
to the BL Lac object 1ES 0502+675. It shows 
absorption lines typical of intermediate spectral type (likely F or G)
stars, not a featureless BL Lac continuum.
}
\label{fig4}
\end{figure}

\begin{figure}
\plotone{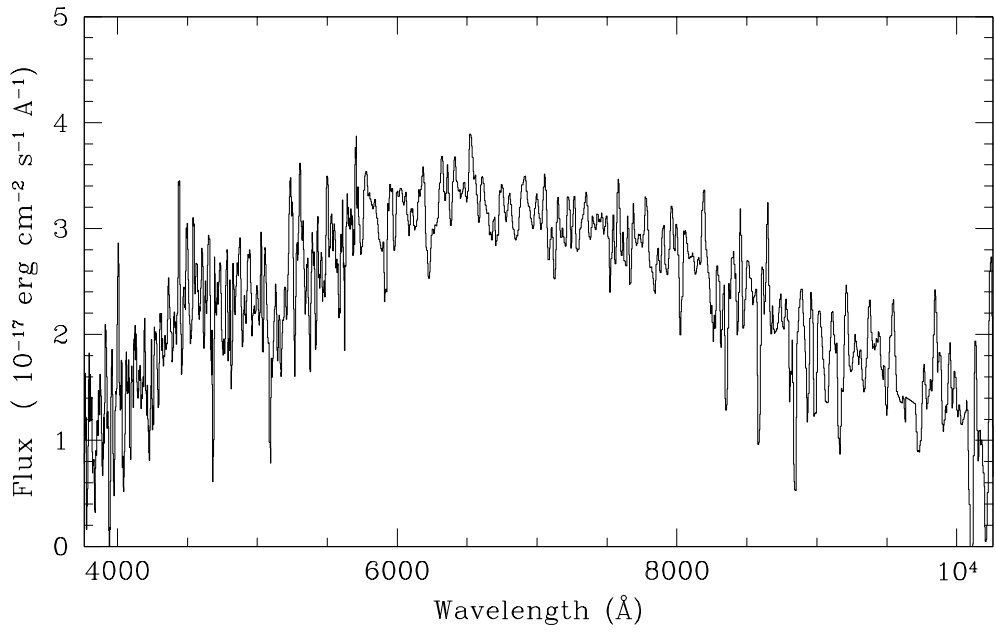}
\caption{STIS optical spectrum of the close companion 
to the BL Lac object 1ES 1440+122. It shows a curved continuum 
suggesting a late
spectral type star (likely K), not a featureless BL Lac power-law continuum.
}
\label{fig5}
\end{figure}

\clearpage

\begin{table} [t]
\begin{center}
\caption{VLA Observations of 1ES 0502+675}
\medskip
\begin{tabular}{ccccc}
\hline \hline
Frequency & HPBW (PA)                 & noise   & Peak Flux   & Total Flux  \\
   GHz    & $^{\prime\prime}$ ($^\circ$) &  mJy/b  &    mJy/b    &    mJy      \\
\hline
 22.5     & 0.087$\times$0.066 (12)&  0.7    &   22.5      &   22.5      \\
 15.0     & 0.16$\times$0.10 (2)   &  0.2    &   24.1      &   24.1      \\
  8.46 A  & 0.31$\times$0.23 (-3)  &  0.03   &   32.00     &   32.10     \\
  8.46 B  & 1.0 $\times$0.8(12)      &  0.013  &   30.4      &   32.0      \\
  8.46 A+B& 0.7$\times$0.5(4)      &  0.01   &   31.2      &   32.0      \\
  1.4     &    45                  &  0.5    &   26.2      &   26.2      \\
\hline
\label{0502vlamaps}
\end{tabular}
\tablecomments{A: VLA A configuration; B: VLA B configuration; 
A+B: Combined data. 1.4~GHz data are from the NVSS survey.}
\end{center}
\end{table}

\begin{table} [t]
\begin{center}
\caption{VLA Observations of 1ES 1440+122}
\medskip
\begin{tabular}{ccccc}
\hline \hline
Frequency & HPBW (PA)                 & noise   & Peak Flux   & Total Flux  \\
   GHz    & $^{\prime\prime}$ ($^\circ$) &  mJy/b  &    mJy/b    &    mJy      \\
\hline
 22.5     & 0.080$\times$0.074 (12)&  0.5    &   20.8      &   20.8      \\
 15.0     & 0.12$\times$0.11 (-1)  &  0.2    &   23.0      &   23.5      \\
  8.46 A  &   0.15                 &  0.05   &   24.4      &   33.0      \\
  8.46 B  & 0.69$\times$0.65 (-3)  &  0.015  &   25.0      &   36.8      \\
  8.46 A+B& 0.61$\times$0.59(4)    &  0.01   &   24.4      &   36.8      \\
  1.4     &    45                  &  0.5    &   66.5      &   70.3      \\
\hline
\label{1440vlamaps}
\end{tabular}
\tablecomments{A: VLA A configuration; B: VLA B configuration; 
A+B: Combined data. 1.4~GHz data are from the NVSS survey.}
\end{center}
\end{table}

\vfill\eject

\begin{table} [t]
\begin{center}
\caption{Measured Positions}
\medskip
\begin{tabular}{lllc}
\hline \hline

Band & R.A. (J2000) & Dec. (J2000)                          & Note \\
     & ~h~  m~~~  s~     &  ~~~$^\circ$~~~ $^\prime$~~~ $^{\prime\prime}$ & \\
\hline

optical & 05 07 56.06 & +67 37 24.0 & star \\
optical & 05 07 56.12 & +67 37 24.1 & BL-Lac \\
radio & 05 07 56.11 & +67 37 24.1 & AGN \\
 
\hline

optical & 14 42 48.27 & +12 00 40.2 & star \\
optical & 14 42 48.28 & +12 00 40.5 & BL-Lac \\ 
radio   & 14 42 48.28 & +12 00 40.4  & AGN \\

\hline
\end{tabular}
\end{center}
\end{table}


\begin{thebibliography}

\bibitem[Abraham et al.(1993)]{ab93}Abraham R. C., Crawford C. S., 
Merrifield M. R., 
Hutchings J. B., McHardy I. M. 1993, \apj, 415, 101.

\bibitem[Condon et al.(1998)]{co98}Condon J.J., Cotton W.D., Greisen E.W., 
Yin Q.F., Perley R.A., 
Taylor G.B., and Broderick J.J.: 1998  \aj, 115. 1693

\bibitem[Falomo et al.(1992)]{fa92}Falomo R., Melnick J., and Tanzi E.G. 
1992, \aap, 255, L17

\bibitem[Falomo(1996)]{fa96}Falomo,R., 1996, \mnras, 283, 241 
 
\bibitem[Falomo et al.(1997)]{fa97}Falomo R., Kotilainen J., 
Pursimo T., et al. 1997, \aap, 321, 374  

\bibitem[Falomo et al.(2000)]{fa00}Falomo R., Scarpa R., Treves A., 
Urry C. M. 2000, \apj, 542, 731

\bibitem[Falomo et al.(2002)]{fa02}Falomo R., Kotilainen J. K., 
Treves A. 2002, \apj, 569, L35 

\bibitem[Giovannini et al.(2001)]{gi01}Giovannini G., Cotton W.D., 
Feretti L., Lara L., Venturi T. 
2001 \apj, 552, 508

\bibitem[Heidt et al.(2003)]{he03}Heidt J., J\"ager K., Nilsson K., 
Hopp U., Fried J.W., 
Sutorius E. 2003 \aap, 406, 565

\bibitem[Lara et al.(2004)]{la04}Lara L., Giovannini G., Cotton W.D., 
Feretti L., Marcaide J.M., Marquez I., Venturi T. 2004 \aap, in press
(astro-ph/0404373)

\bibitem[Lewis \& Ibata(2000)]{le00}Lewis G.F. \& Ibata R.A. 2000 \apj, 528 650
 
\bibitem[Maoz et al.(1993)]{ma93}Maoz D., Bahcall J.N., Schneider D.P. 
et al. 1993, \apj, 409, 28
 
\bibitem[Nilsson et al.(1996)]{ni96}Nilsson K., Charles P.A., 
Pursimo T., Takalo L. O., 
Sillanp\"a\"a A., Teerikorpi, P.
        1996 \aap, 314 754 

\bibitem[Ostriker \& Vietri(1985)]{os85}Ostriker J.P. \& Vietri M. 1985, 
\nat, 318, 446

\bibitem[Ostriker \& Vietri(1990)]{os90}Ostriker J.P. \& Vietri M. 1990, 
\nat, 344, 45

\bibitem[Pian et al.(2002)]{pi02}Pian E. \& al. 2002 \aap, 392 407

\bibitem[Pursimo et al.(2002)]{pu02}Pursimo T., Nilsson K., Takalo L.O., 
Sillanp\"a\"a S., Heidt J. and Pietila H.
        2002 \aap, 381 810

\bibitem[Rector \& Stocke(2003)]{re03}Rector T. \& Stocke J.T.  
2003 \aj, 125 2447

\bibitem[de Ruiter et al.(1990)]{ru90}de Ruiter H.R., Parma P., Fanti C., 
Fanti R. 1990 \aap, 227, 351

\bibitem[Scarpa et al.(1999)]{sc99}Scarpa R., Urry C.M., Falomo R., 
Pesce J.E., Webster R., 
O'Dowd M., Treves A. 1999 \apj, 521, 134

\bibitem[Scarpa et al.(2000)]{sc00}Scarpa R., Urry C.M., Falomo R., 
Pesce J.E., Treves A. 
2000, \apj, 532, 740

\bibitem[Stickel et al.(1988a)]{st88a}Stickel M., Fried J.W. \& K\"uhr,H. 
1988a, \aap, 198, L13

\bibitem[Stickel et al.(1988b)]{st88b}Stickel M., Fried J.W. \& K\"uhr,H. 
1988b, \aap, 206, L30
 
\bibitem[Stocke et al.(1995)]{st95}Stocke J.T., Wurtz R.E., Perlman E.S. 
1995, \apj, 454, 55
 
\bibitem[Stocke \& Rector(1997)]{st97}Stocke J.T. \& Rector, T. 1997 \apj, 
489 L17

\bibitem[Urry \& Padovani(1995)]{ur95}Urry C.M. \& Padovani, P. 1995, 
\pasp, 107, 803

\bibitem[Urry et al.(2000)]{ur00} Urry C.M., Scarpa R. O'Dowd, M. 
Falomo R., Pesce J., Treves A. 2000, \apj, 532, 816

\end{thebibliography}
\end{document}